\documentclass[conference]{IEEEtran}
\IEEEoverridecommandlockouts
\usepackage{cite}
\usepackage{amsmath,amssymb,amsfonts}
\usepackage{algorithmic}
\usepackage{graphicx}
\usepackage{textcomp}
\usepackage{xcolor}

\usepackage{algorithm}
\usepackage{url}
\usepackage{hyperref}
\usepackage[caption=false,font=normalsize,labelfont=sf,textfont=sf]{subfig}
\usepackage{soul}
\usepackage{multirow}
\usepackage{svg}

\usepackage{balance}
\usepackage{comment}

\def\BibTeX{{\rm B\kern-.05em{\sc i\kern-.025em b}\kern-.08em
    T\kern-.1667em\lower.7ex\hbox{E}\kern-.125emX}}

\makeatletter
\newcommand{\linebreakand}{%
  \end{@IEEEauthorhalign}%
  \hfill\mbox{}\par
  \mbox{}\hfill\begin{@IEEEauthorhalign}}
\makeatother

\begin{document}

\title{Cross-Domain Knowledge Transfer for Underwater Acoustic
Classification Using Pre-trained Models
%\\
%{\footnotesize \textsuperscript{*}Note: Sub-titles are not captured in %Xplore and should not be used}
\thanks{DISTRIBUTION STATEMENT A. Approved for public release. Distribution is unlimited. This material is based upon work supported by the Under Secretary of Defense for Research and Engineering under Air Force Contract No. FA8702-15-D-0001. Any opinions, findings, conclusions or recommendations expressed in this material are those of the author(s) and do not necessarily reflect the views of the Under Secretary of Defense for Research and Engineering. Code is publicly available at this Github \url{https://github.com/Advanced-Vision-and-Learning-Lab/PANN_Models_DeepShip}.}
}

\begin{comment}
\newcommand{\linebreakand}{%
  \end{@IEEEauthorhalign}
  \hfill\mbox{}\par
  \mbox{}\hfill\begin{@IEEEauthorhalign}
}

\author{
\IEEEauthorblockN{1\textsuperscript{st} Amirmohammad Mohammadi}
\IEEEauthorblockA{\textit{Department of Electrical and Computer Engineering} \\
\textit{Texas A\&M University}\\
College Station, TX, USA \\
amir.m@tamu.edu}
\and
\IEEEauthorblockN{2\textsuperscript{nd} Tejashri Kelhe}
\IEEEauthorblockA{\textit{Department of Electrical and Computer Engineering} \\
\textit{Texas A\&M University}\\
College Station, TX, USA \\
tkelhe@tamu.edu}
\linebreakand
\IEEEauthorblockN{3\textsuperscript{rd} Davelle Carreiro}
\IEEEauthorblockA{\textit{Massachusetts Institute of Technology Lincoln Laboratory} \\
Lexington, MA, USA \\
davelle.carreiro@ll.mit.edu}
\and
\IEEEauthorblockN{4\textsuperscript{th} Alexandra Van Dine}
\IEEEauthorblockA{\textit{Massachusetts Institute of Technology Lincoln Laboratory} \\
Lexington, MA, USA \\
alexandra.vandine@ll.mit.edu}
\linebreakand
\IEEEauthorblockN{5\textsuperscript{th} Joshua Peeples}
\IEEEauthorblockA{\textit{Department of Electrical and Computer Engineering} \\
\textit{Texas A\&M University}\\
College Station, TX, USA \\
jpeeples@tamu.edu}
}
\end{comment}

\author{
  % --- Row 1: first two authors ---
  \IEEEauthorblockN{1\textsuperscript{st} Amirmohammad Mohammadi}%
  \IEEEauthorblockA{%
    \textit{Department of Electrical and Computer Engineering}\\
    \textit{Texas A\&M University}\\
    College Station, TX, USA\\
    amir.m@tamu.edu%
  }%
  \and
  \IEEEauthorblockN{2\textsuperscript{nd} Tejashri Kelhe}%
  \IEEEauthorblockA{%
    \textit{Department of Electrical and Computer Engineering}\\
    \textit{Texas A\&M University}\\
    College Station, TX, USA\\
    tkelhe@tamu.edu%
  }%
  \linebreakand
  % --- Row 2: next two authors ---
  \IEEEauthorblockN{3\textsuperscript{rd} Davelle Carreiro}%
  \IEEEauthorblockA{%
    \textit{Massachusetts Institute of Technology Lincoln Laboratory}\\
    Lexington, MA, USA\\
    davelle.carreiro@ll.mit.edu%
  }%
  \and
  \IEEEauthorblockN{4\textsuperscript{th} Alexandra Van Dine}%
  \IEEEauthorblockA{%
    \textit{Massachusetts Institute of Technology Lincoln Laboratory}\\
    Lexington, MA, USA\\
    alexandra.vandine@ll.mit.edu%
  }%
  \linebreakand
  % --- Row 3: single centered fifth author ---
  \IEEEauthorblockN{\hfill 5\textsuperscript{th} Joshua Peeples\hfill}%
  \IEEEauthorblockA{\hfill
    \textit{Department of Electrical and Computer Engineering}\\
    \textit{Texas A\&M University}\\
    College Station, TX, USA\\
    jpeeples@tamu.edu}%
}

\maketitle

\begin{abstract}
Transfer learning is commonly employed to leverage large, pre-trained models and perform fine-tuning for downstream tasks. The most prevalent pre-trained models are initially trained using ImageNet. However, their ability to generalize can vary across different data modalities. This study compares pre-trained Audio Neural Networks (PANNs) and ImageNet pre-trained models within the context of underwater acoustic target recognition (UATR). It was observed that the ImageNet pre-trained models slightly out-perform pre-trained audio models in passive sonar classification. We also analyzed the impact of audio sampling rates for model pre-training and fine-tuning. This study contributes to transfer learning applications of UATR, illustrating the potential of pre-trained models to address limitations caused by scarce, labeled data in the UATR domain.
\end{abstract}


\section{Introduction}
\label{sec:intro}

Deep learning is used frequently for audio classification tasks \cite{Survey2023} due to its ability to automatically extract relevant data features and identify complex patterns. These tasks have broad applications from healthcare\cite{esposito2022quantum}, urban development \cite{tyagi2023urban}, and environmental monitoring \cite{mu2021environmental} to underwater acoustic target recognition (UATR) \cite{tian2021deep}. In a subset of the latter application family, passive sonar can be used to identify objects at the water surface or underwater using knowledge of target acoustics signatures and expected propagation through the water column. This capability impacts numerous maritime tasks including analysis of biological life cycles\cite{thomas2020marine}, assisted search and rescue operations \cite{10309055}, monitor of maritime traffic \cite{beckler2022multilabel}, and analysis of unknown sound sources \cite{yang2020underwater}.

Obtaining large, labeled, publicly available datasets for UATR is difficult \cite{yao2024underwater}. To overcome this challenge, transfer learning can be used to leverage large, pre-trained models \cite{li2020transfer}. Transfer learning has several advantages including data efficiency, shorter training time, and increased performance \cite{zhuang2020comprehensive}. Our work leverages pre-trained Audio Neural Networks (PANNs) \cite{kong2020panns} and ImageNet pre-trained models via PyTorch Image Models (TIMMs)\cite{rw2019timm}, to the domain of UATR using a publicly available dataset, DeepShip \cite{irfan2021deepship}.

PANN models are trained on the large-scale AudioSet dataset \cite{gemmeke2017audio}, which has over 5,000 hours of audio recordings from YouTube videos. These models are proposed to be transferable to other audio pattern recognition tasks. The TIMM library includes a number amount of models, many of which are trained on ImageNet1K \cite{ImageNet}. PANNs and TIMMs aim to leverage knowledge gained from large datasets and transfer knowledge to other, seemingly unrelated, tasks. Our work aims to shed light on the usefulness of different types of pre-trained models for passive sonar classification. This study is also the first to thoroughly investigate PANNs for the DeepShip dataset. Related work \cite{yao2024underwater,wu2024recognizing} used PANNs for a smaller passive sonar dataset (ShipsEar \cite{santos2016shipsear}).

\section{Method}
\label{sec:method}

\subsection{Model Preparation}
\label{ssec:subheadm2}

The PANN CNN14 model \cite{kong2020panns} was used for the first portion of investigation since this model was pre-trained across three different data sampling rates of 8, 16, and 32 kHz on the AudioSet dataset. We investigated the performance of these pre-trained models on DeepShip \cite{irfan2021deepship} data sampled at the different rates in order to determine if the sampling frequency of a) data and b) the pre-trained model impacts classification performance. The second portion of investigation focused on the five top-performing PANN models (CNN14, ResNet38, MobileNetV1, Res1dNet31, and Wavegram-Logmel-CNN) and a number of pre-trained ImageNet1K models from TIMM (ResNet50, DenseNet201, MobilenetV3-large-100, ConvNeXtV2-tiny, and RegNety-320). TIMM models were chosen to be either equivalent to their PANN counterparts (\textit{e.g.}, MobileNets and ResNets) or were recently published (\textit{i.e.,} ConvNeXtV2-tiny).

\begin{figure*}[t]
\centering
\subfloat[]{\includegraphics[width=0.35\textwidth]{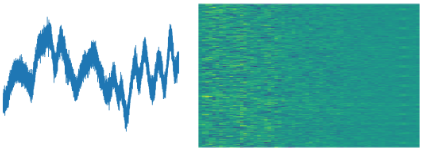}%
\label{fig:preprocess}}
\hfil
\subfloat[]{\includegraphics[width=0.194\textwidth]{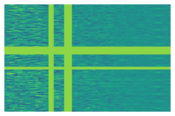}%
\label{fig:augmentation}}
\hfil
\subfloat[]{\includegraphics[width=0.39\textwidth]{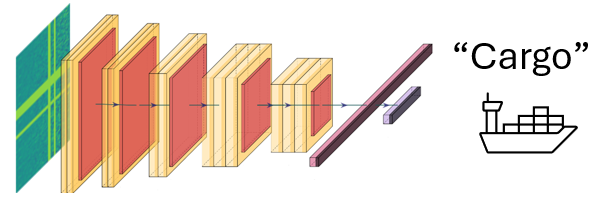}%
\label{fig:analysis}}
\caption{The overall framework is shown. (a) Data Preprocessing: The audio waveform is transformed into a logarithmic mel-frequency spectrogram. (b) Data Augmentation: SpectAugmentation \cite{park2019specaugment} and Mixup \cite{zhang2018mixup} are added during training to improve model robustness. (c) Data Analysis: Input spectrograms are processed using networks pre-trained on AudioSet (PANN \cite{kong2020panns}) or ImageNet (TIMM \cite{rw2019timm}).}
\label{fig:framework}
\end{figure*}

To adapt both PANN and TIMM models for the UATR task, full fine-tuning of the models was performed. As noted by prior work \cite{kong2020panns}, full fine-tuning maximizes performance as opposed to training from scratch or freezing portions of the networks. For both PANN and TIMM models, the waveform signals were first converted to spectrograms before being passed into the models. 

\subsection{Data Preparation}
\label{ssec:subheadm1}

The DeepShip dataset includes four classes of different ship types labeled as cargo, passengership, tanker, and tug. To investigate the impact of frequency resolution on model performance, the signals were resampled to frequencies of 8 kHz, 16 kHz, and 32 kHz and then segmented into intervals of five seconds where each of these segments is an individual example for the training paradigm. Three datasets were, therefore, generated based on sampling rate where each contained a total of 609 recordings and 33,770 segments.

The dataset was partitioned into training, validation, and test sets with ratios of 70\%, 10\%, and 20\%, respectively, using a method similar to that described in \cite{ritu2023histogram}, which splits the dataset based on recordings. To prevent data leakage, if a specific recording segment was chosen for training, all other segments in that recording were also chosen for training. The dataset split was completed using stratification to maintain similar distribution of all four classes across the sets. To reproduce these partitions across experiments, the dataset split was initially written and saved in a file that was used to generate subsequent data partitions. Specifically, the dataset was split into training, validation, and testing sets with 428 recordings (23,088 segments), 60 recordings (3,974 segments), and 121 recordings (6,708 segments), respectively. Normalization was applied to each data split using a minimum-maximum approach, i.e., by computing the global minimum and maximum values from the training dataset.

Each segment was converted into a spectrogram using the Short-Time Fourier Transform (STFT) with a Hann window of size 1024 and a hop length of 320, a parameterization which captures the frequency content of the signal over time. The resulting spectrogram is then transformed into a logarithmic mel-frequency spectrogram by applying a logarithmic mel-frequency filter bank with 64 mel filters, which compresses the frequency information into the mel-frequency scale. The lower and upper cut-off frequencies for the mel-frequency filters are set to 50 Hz and 14 kHz, respectively, to filter out low-frequency noise and minimize aliasing effects. 

\subsection{Data Augmentation}
\label{sect:data_aug}
Following \cite{kong2020panns}, to improve model robustness to variations in the data, SpecAugmentation \cite{park2019specaugment} is applied on the spectrogram, a technique which incorporates transformations such as time and frequency masking. Mixup \cite{zhang2018mixup} is applied to the spectrogram to further augment the data by mixing the inputs and targets of two samples, thereby incorporating adversarial training examples. Each data augmentation technique was used for both PANN and TIMM models for a fair comparison. Because TIMM models are initially trained on 3-channel (RGB) ImageNet data, when using TIMM models for single-channel inputs (\textit{i.e.}, spectrograms), the weights of the first convolutional layer are summed across the three input kernels. Aggregation of the weights allows the pre-trained TIMM models to be applied to single channel inputs without needing to modify the remaining portion of the model's structure. PANN models, in contrast, are able to be applied directly to the passive sonar data as the pre-training was accomplished with single channel data features.

\section{Experimental Procedure}
\label{sec:experiments}

For the first experimental investigation, each CNN14 model was fine-tuned, pre-trained at three distinct sampling rates, and resampled the DeepShip dataset to match these frequencies. This resulted in the evaluation of nine data-model combinations. For the CNN14 models pre-trained on 16 kHz and 8 kHz, the window length, hop size, and upper cut-off frequency were adjusted to half and one-quarter, respectively, relative to the 32 kHz model. This scaling was necessary to maintain consistent resolution across different sampling rates, so that each model would ingest spectrograms appropriate to the model configuration.

To make data augmentation consistent across different data sampling rates, the time mask width for SpecAugmentation was adjusted based on the ratio of the data sample rate to the model sample rate. The formula used is given by
(\ref{eqn:mask}):

\begin{equation}
W_{\text{mask}} = W_{\text{base}} \times \left( \frac{r_{\text{data}}}{r_{\text{model}}} \right) \label{eqn:mask}
\end{equation}

\noindent where $W_{\text{base}}$ is set to 64. For example, when both the model sample rate ($r_{\text{model}}$) and the data sample rate ($r_{\text{data}}$) are 32 kHz, the time mask width remains 64. However, if the data sample rate is 16 kHz whereas the model sample rate is 32 kHz, the time mask width is adjusted to 32.

As a result, the number of time frames are also halved (from 501 to 251) in this example. This ensures appropriate scaling between augmentation schema and the data sample rate. The comparison of pre-trained audio (PANN) and image (TIMM) models was performed using a fixed audio frequency resolution of 32 kHz as all available PANN models (except CNN14) were pre-trained on 32 kHz data. 

All models were trained using a learning rate of 5e-5, a batch size of 64, and the Adam optimizer for 100 epochs with a patience setting of 50 epochs based on validation loss. The batch size was set to 32 for the impact of sampling rate experiments in Section \mbox{\ref{ssec:subheadR1}}. All experiments were completed on an NVIDIA A40 GPU with each experiment conducted over three random runs of initialization to evaluate the reproducibility of the results. The overall framework is depicted in Fig. \ref{fig:framework}. Models are then fine-tuned on this data for classification. During training, data augmentation is incorporated as described in Section \ref{sect:data_aug}.

\section{Results and Discussion}
\label{sec:results_discussion}

\subsection{Impact of Sampling Rate}
\label{ssec:subheadR1}

\begin{figure}[htb]
    \centering
    %\includesvg[width=0.95\linewidth]{Figures/freq_accuracy_plot_v4.svg}
    \includegraphics[width=0.95\linewidth]{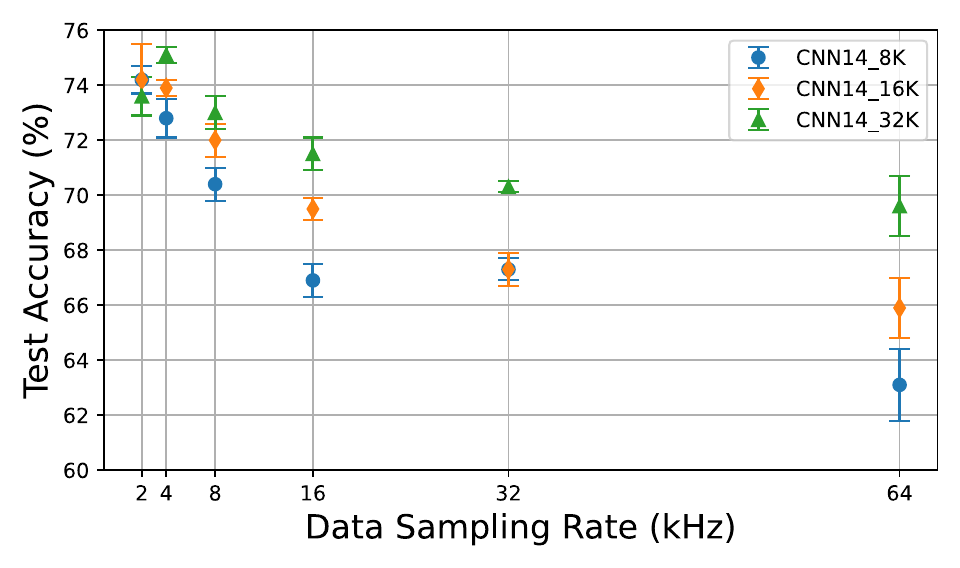}

    \caption{Average test accuracy across three experimental runs at different sampling rates with $\pm1$ standard deviation for CNN14 models. Each color represents a different model, with the symbol of each model centered on the average test accuracy and the error bars show $\pm1$ standard deviation.}
    \label{fig:freq_accuracy_plot}
\end{figure}

\begin{table}[ht]
    \centering
    \caption{Average Test Accuracy Across Three Experimental Runs at Different Sampling Rates With $\pm1$ Standard Deviation. The Highest Average Test Accuracy Is Bolded.}
    \label{tab:accuracy}
\begin{tabular}{|c|c|c|c|}
\hline
\multirow{2}{*}{Data Sampling Rate} & \multicolumn{3}{c|}{Models} \\ \cline{2-4} 
                        & CNN14\_8K              & CNN14\_16K             & CNN14\_32K              \\ \hline
2 kHz                   & $74.2 \pm 0.5\%$        & $74.2 \pm 1.3\%$        & $73.6 \pm 0.7\%$         \\ \hline
4 kHz                   & $72.8 \pm 0.7\%$        & $73.9 \pm 0.3\%$        & $\mathbf{75.1 \pm 0.3\%}$         \\ \hline
8 kHz                   & $70.4 \pm 0.6\%$       & $72.0 \pm 0.6\%$       & $73.0 \pm 0.6\%$ \\ \hline
16 kHz                  & $66.9 \pm 0.6\%$       & $69.5 \pm 0.4\%$       & $71.5 \pm 0.6\%$        \\ \hline
32 kHz                  & $67.3 \pm 0.4\%$       & $67.3 \pm 0.6\%$       & $70.3 \pm 0.2\%$        \\ \hline
64 kHz                   & $63.1 \pm 1.3\%$        & $65.9 \pm 1.1\%$        & $69.6 \pm 1.1\%$         \\ \hline
\end{tabular}
\end{table}

The impact of different frequency resolutions on the performance of various CNN14 models is summarized in Table \ref{tab:accuracy}. Results show that the CNN14\_32k model achieves the highest accuracy on held-out test data sampled at 4 kHz with $75.1 \pm 0.3\%$. Fig. \ref{fig:freq_accuracy_plot} illustrates the average test accuracy across different data sampling rates. The error bars represent the standard deviation across three experimental runs. In this visual, clear trends can be observed over the various data sampling rates. The CNN14\_32K model was robust across the different sampling rates as opposed to the other lower sampling rate models.

These results suggest that the pre-trained models  perform better when tested over lower data sample rates, while higher data sample rates do not necessarily improve accuracy. One advantage of a lower data sampling rate is reduced computational cost since spectrogram features have smaller sizes which require less memory. Also, all three models perform comparably when the data sampling rate is low. This finding suggests that higher frequency resolution training does not necessarily decrease a model's ability to generalize to lower frequency resolutions. Rather, it appears that models are able to extract useful features even with reduced data resolution during fine-tuning.
The 32 kHz model was pre-trained on data with finer details (32,000 samples per second) compared to the 8 and 16 kHz models. Coarser DeepShip samples are still able to be correctly classified using the knowledge gained from pre-training on the Audioset dataset \cite{gemmeke2017audio}. Therefore, models pre-trained with higher frequency resolution data can be robust to fine-tuning data with different sampling rates, a finding which benefits applications where alternate sampling rates are used to process data as is the case with many UATR datasets.

\subsection{Impact of Pre-training Data}
\label{ssec:subheadR2}

\begin{table}[ht]
    \centering
    \caption{Average Test Accuracy for PANN and TIMM Models Across Three Experimental Runs of Random Initialization With $\pm1$ Standard Deviation. The Highest Average Test Accuracy Is Bolded.}
    \label{tab:model_comparison}
    \begin{tabular}{|c|c|c|c|}
        \hline
        Type & Model Name & Accuracy & Parameters \\ \hline
        \multirow{5}{*}{PANN} & CNN14-32k & $70.6 \pm 0.8\%$ & 79.7M \\ 
         & Wavegram-Logmel-CNN & $68.6 \pm 2.7\%$ & 80.0M \\
         & MobileNetV1 & $67.1 \pm 0.3\%$ & 4.3M \\ 
         & ResNet38 & $66.3 \pm 0.3\%$ & 72.7M \\ 
         & Res1dNet31 & $63.2 \pm 1.8\%$ & 79.4M \\ \hline
        \multirow{5}{*}{TIMM} & ConvNeXtV2-tiny & $\mathbf{73.7 \pm 0.8\%}$ & 27.9M \\ 
         & RegNety-320 & $72.5 \pm 1.1\%$ & 141M \\ 
         & DenseNet201 & $68.5 \pm 2.0\%$ & 18.1M \\ 
         & ResNet50 & $68.2 \pm 0.7\%$ & 23.5M \\ 
         & MobilenetV3-large-100 & $65.4 \pm 0.7\%$ & 4.2M \\ \hline
    \end{tabular}
\end{table}

The comparison between PANN and TIMM models is presented in Table \ref{tab:model_comparison}. Within the PANN models, the CNN14-32k shows the highest accuracy at 70.6 $\pm$ 0.8\%, suggesting that this model is better suited to the task of passive sonar classification. This result is also consistent with that shown in the PANN \cite{kong2020panns} work asserting that CNN14 was the best model across a variety of audio datasets. The PANN Wavegram-Logmel-CNN also performs competitively, considering the larger error margin. Amongst the TIMM models, the ConvNeXtV2-tiny model stands out with the highest test accuracy of 73.7 $\pm$ 0.8\%. RegNety-320 also performs competitively, however with five times the number of model parameters. The results show that more complex models (\textit{i.e.}, more parameters) do not necessarily lead to a proportional increase in accuracy. The ConvNeXTV2-tiny model's self-supervised training approach which uses a fully convolutional masked autoencoder may provide some explanation as to its high comparative performance \cite{woo2023convnext} .

Despite having a lower number of parameters, the MobileNetV1 still performs competitively with more computationally expensive models. The model's efficient use of parameters could be useful for applications requiring lightweight architectures without sacrificing accuracy. When compared to the pre-trained ImageNet version (MobilenetV3-large-100), there is a statistically significant difference in performance. This result is intuitive as one would expect that PANN models should out-perform models pre-trained on image data when fine-tuning on sonar data as it is a much closer analog to audio data than photo imagery. However, we do not see this trend overall as most of the PANN models are outperformed by the TIMM models. Pre-training on a large dataset such as ImageNet appears to capture features that can be broadly useful in other data modalities (such as passive sonar).

\subsection{Best PANN and TIMM Models Comparison}

\begin{figure}[htb]
\centering
\subfloat[]{\includegraphics[width=0.235\textwidth]{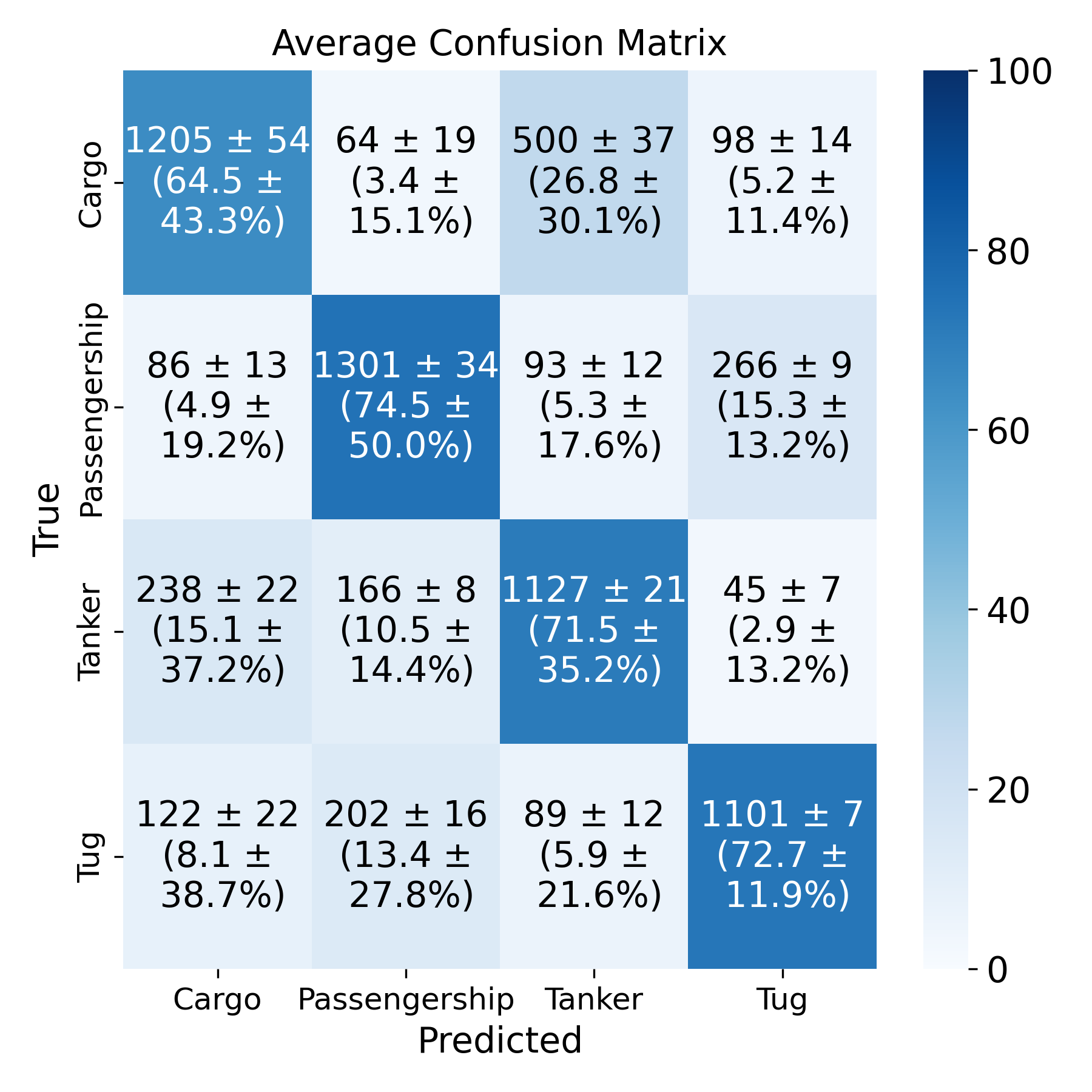}%
\label{fig:confusion_matrix_pann}}
\hfil
\subfloat[]{\includegraphics[width=0.235\textwidth]{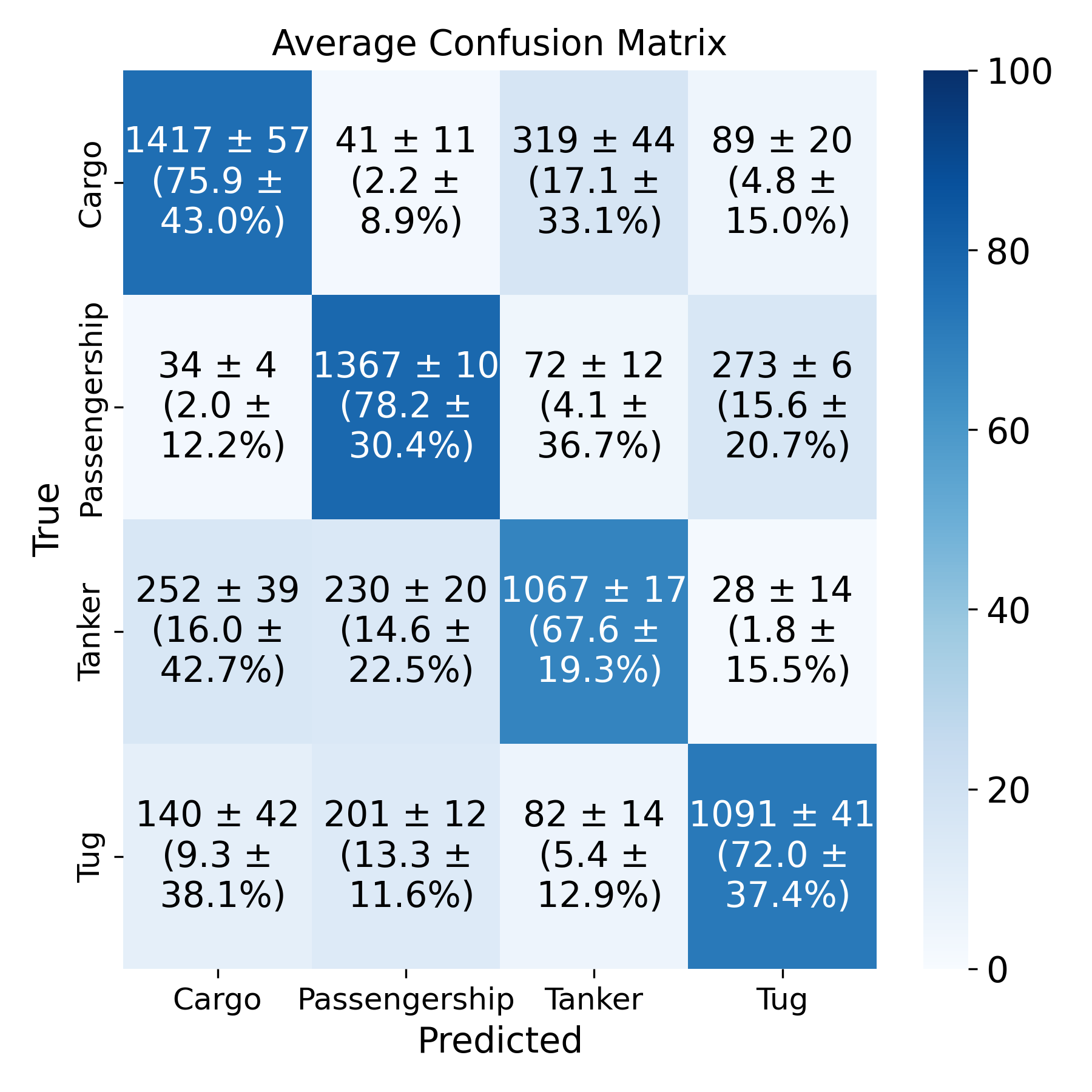}%
\label{fig:confusion_matrix_timm}}
\caption{Average confusion matrices for best PANN model (a) CNN14-32k (70.6 $\pm$ 0.8) and best TIMM model (b) ConvNeXtV2-tiny (73.7 $\pm$ 0.8) across three experimental runs. The average test accuracy $\pm1$ standard deviation is shown in parentheses.}
\label{fig:confusion_matrices}
\end{figure}

To further analyze the performance of the models, confusion matrix was referenced. This visualization provides deeper insights into the model performance characteristics beyond accuracy metrics. According to the confusion matrices in Fig. \ref{fig:confusion_matrices} illustrating the best PANN and TIMM models, the CNN14 model better identifies the tanker class when compared to the ConvNeXtV2-tiny model on average. This result indicates that the CNN14 has learned distinctive features specific to this class, which can be more useful in applications where precise classification of vessel types is required. It is worth noting the lower standard deviation for ConvNeXtV2-tiny, suggesting a more consistent performance across different trials, and therefore, potentially a more robust model. Despite ConvNeXtV2-tiny outperforming other models, the model had difficulty differentiating between cargo and tanker as well as between passengership and tug classes. This result suggests that the learned features are not distinct enough. Additional spectrogam features or other learning paradigms (\textit{e.g.}, contrastive learning) may help capture unique acoustic signatures of each class.

{Also, Grad-CAM \mbox{\cite{selvaraju2017grad}} was leveraged to visualize class-discriminative regions in the learned feature representations. Specifically, per-sample class activation maps were extracted from the final convolutional layer of both best models gradient weighted activations computed, and the resulting heatmaps normalized. CAMs were then aggregated across all correctly classified and misclassified samples within each class. This enabled assessment of how the two architectures focus their attention on the spectro-temporal patterns of the input log-mel spectrograms.}

\begin{figure*}[htb]
\centering

% First Row
\subfloat[]{%
\includegraphics[width=0.49\textwidth]{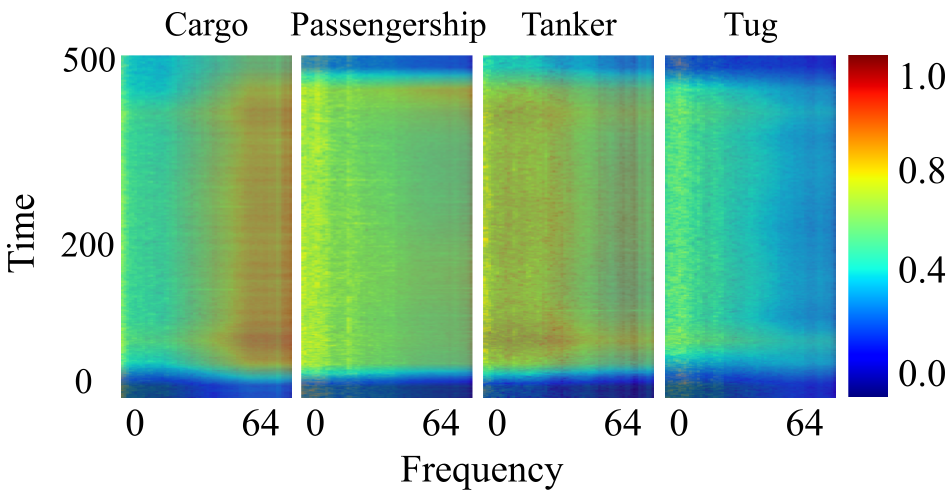}%
\label{fig:grad_cam_pann_correct}}
\hfil
\subfloat[]{%
\includegraphics[width=0.49\textwidth]{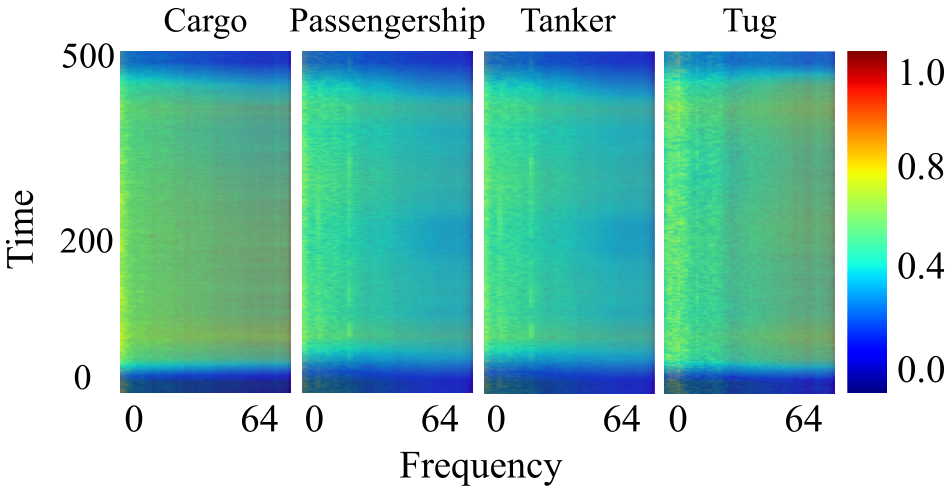}%
\label{fig:grad_cam_pann_misclassified}}
\\
% Second Row
\subfloat[]{%
\includegraphics[width=0.49\textwidth]{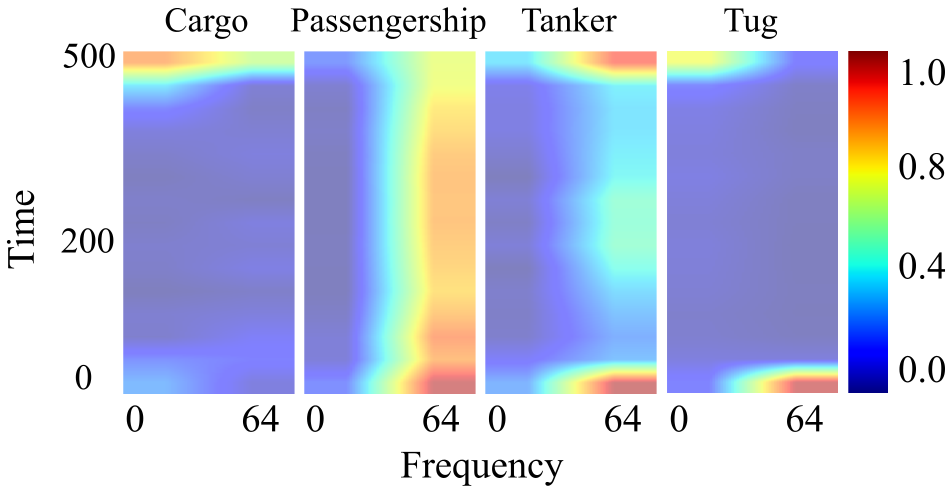}%
\label{fig:grad_cam_timm_correct}}
\hfil
\subfloat[]{%
\includegraphics[width=0.49\textwidth]{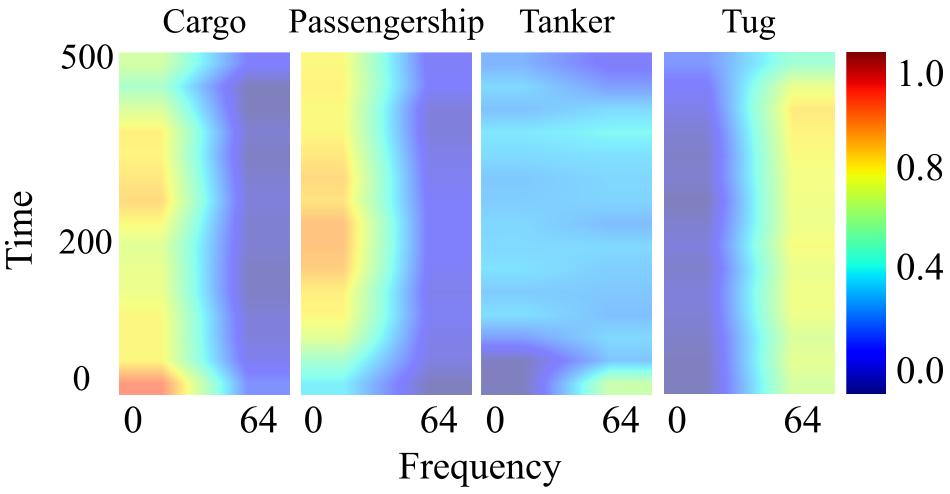}%
\label{fig:grad_cam_timm_misclassified}}

\caption{Visualization of Grad-CAM results for (a) CNN14 correctly classified samples, (b) CNN14 misclassified samples, (c) ConvNeXtV2-tiny correctly classified samples, and (d) ConvNeXtV2-tiny misclassified samples.}
\label{fig:grad_cam_grid}
\end{figure*}

{As shown in Fig. \mbox{\ref{fig:grad_cam_grid}}, for correctly classified samples, the ConvNeXtV2-tiny displayed a more narrowly focused activation on specific regions, while the CNN14 exhibited broader activation patterns across the spectrogram. Conversely, for misclassified samples, the TIMM model shifted towards a wider attention span, while the PANN model continued to show broad activation, suggesting less discriminative refinement. In conclusion, these observations suggest that the best TIMM model could be the more interpretable choice, given a clear attention shift dependent on prediction correctness, potentially aiding in model explainability and refinement.}

\section{Conclusion}
\label{sec:conc}

This investigation into the application of pre-trained models to passive sonar classification using the DeepShip dataset contributes to community understanding of the influence signal processing parameterization and model pre-training can have on classifier performance. It was confirmed that adjusting the data sampling rate can impact the performance of neural networks, with the CNN14-32k model displaying higher performance across lower frequencies. Furthermore, our comparative analysis of PANN and TIMM models showed that models trained on visual datasets slightly out-performed those trained on audio datasets when fine-tuned and applied to passive sonar classification tasks. Results, herein, show the importance of exploring architectures based on their performance across data modalities, rather than strictly applying models on the same data used for pre-training. 

Future work could explore the integration of multi-modal data to further improve the accuracy of classification models in complex acoustic environments. Additionally, self-supervised learning approaches such as masked autoencoders \cite{woo2023convnext} may assist in more effectively modifying the feature representation of all models for UATR. Other interesting areas of study include investigating parameter efficient transfer learning approaches such as adapters \cite{houlsby2019parameter} instead of fully fine-tuning the models in order to limit computational expense. Given their widely touted state-of-the-art capabilities, comparison to transformer architectures as opposed to convolution neural networks would also be an interesting research topic to explore.

\section*{Acknowledgment}
Portions of this research were conducted with the advanced computing resources provided by Texas A\&M High Performance Research Computing. 

\balance
\bibliographystyle{IEEEtran}
\bibliography{refs}

%\begin{thebibliography}{00}
%\bibitem{b1} G. Eason, B. Noble, and I. N. Sneddon, ``On certain integrals %of Lipschitz-Hankel type involving products of Bessel functions,'' Phil. %Trans. Roy. Soc. London, vol. A247, pp. 529--551, April 1955.
%\end{thebibliography}

\end{document}